\documentclass[12pt]{iopart}
\usepackage{amssymb}

\usepackage{graphicx}
\usepackage{subfigure}

\begin{document}

\title{Far-from-equilibrium transport with constrained resources}
\author{D.A. Adams$^{1,2}$, B. Schmittmann$^{2}$, and R.K.P. Zia$^{2}$}
\date{\today }

\begin{abstract}
\noindent The totally asymmetric simple exclusion process (TASEP) is a well
studied example of far-from-equilibrium dynamics. Here, we consider a TASEP
with open boundaries but impose a global constraint on the total number of
particles. In other words, the boundary reservoirs and the system must share
a finite supply of particles. Using simulations and analytic arguments, we
obtain the average particle density and current of the system, as a function
of the boundary rates and the total number of particles. Our findings are
relevant to biological transport problems if the availability of molecular
motors becomes a rate-limiting factor.
\end{abstract}

\address{$^{1}$Department of Physics, University of Michigan, Ann Arbor, MI 48109, USA \\
$^{2}$Department of Physics, Virginia Tech, Blacksburg, VA 24061, USA} %
\ead{\mailto{davidada@umich.edu}, \mailto{schmittm@vt.edu},
\mailto{rkpzia@vt.edu}}





\section{Introduction}

Over the last two decades, the study of driven diffusive systems \cite%
{KLS,SZ,VP,Schutz} has formed a significant branch in the general pursuit to
understand non-equilibrium statistical mechanics. The underlying dynamics
violates detailed balance, so that the long-time behavior is governed by a
true \emph{non-equilibrium} steady state (NESS). Highly unexpected and
non-trivial properties are manifested, even in one-dimensional systems with
purely local dynamics. A paradigmatic model in the last class is the totally
asymmetric simple exclusion process (TASEP) \cite%
{Krug,Derrida92,DEHP,S1993,Derrida,Schutz}. Characterized by open boundaries
and particle transport, it displays the key physical signatures of a system
driven far from equilibrium. In particular, its NESS develops a complex
phase diagram controlled by the interactions of the system with its
environment. While an understanding of these properties is of fundamental
theoretical interest, the TASEP and its generalizations have also acquired
fame as models for practical problems in traffic flow \cite{Chowdhury,Popkov}
and biological transport \cite{MG,MG2,TC,LBSZia,TomChou,JJ,JJ2,Howard,KF,BC}.

In this paper, we study a simple variant of the basic TASEP, by introducing
a global constraint on the available number of particles. Deferring the
motivations for such a model to the next paragraphs, let us briefly
summarize the essentials. Each site of a one-dimensional lattice is either
empty or occupied by a single particle. Following random sequential
dynamics, the particles enter the lattice at one end (e.g., the ``left''
edge) with a given rate $\alpha $, hop to the right with rate $\gamma $
(scaled to unity), subject to an excluded volume constraint, and exit at the
far end with a rate $\beta $. In the standard version of the model, both $%
\alpha $ and $\beta $ are constant rates, independent of the number of
particles on the lattice. Thus, we may regard the lattice as being coupled
to a reservoir -- or a pool -- with an arbitrarily large number of
particles. Here, we consider a TASEP with a \emph{finite} supply of
particles and report our numerical and analytical findings. A fixed number
of particles, $N_{tot}$, is shared between the lattice and the reservoir, in
such a way that the entrance rate, $\alpha $, depends on the number in the
pool: $N_{p}$. As a result, if more particles are found on the lattice, the
number of available particles in the pool is lowered, leading to a
corresponding decrease of $\alpha $. Since the number of particles on the
TASEP lattice, $N$, feeds back into $\alpha $ through the constraint $%
N+N_{p}=$ constant $=N_{tot}$, we will refer to our model as ``a constrained
TASEP.'' For simplicity, we assume the reservoir to be so large that
particles exiting the lattice are unaffected by $N_{p}$, leaving $\beta $
unchanged. Our focus here is to explore the consequences of limited
resources: How will the phase diagram and the properties of the phases be
affected as the total number of available particles is reduced?

This behavior mimics the limited availability of resources required for a
given physical or biological process. For example, in protein synthesis,
ribosomes bind near the start sequence of a messenger RNA, in order to
translate the genetic information encoded on the RNA into the associated
protein. Ribosomes are large molecular motors which are assembled out of
several basic units, and numerous ribosomes can be bound to the same, or
other, mRNAs, so that multiple proteins are translated in parallel. When a
protein is completed, the ribosome is disassembled and recycled into the
cytoplasm. Under conditions of rapid cell growth, ribosomes or their
constituents can find themselves in short supply, so that a self-limitation
of translation, mimicked via a modification of $\alpha $, can occur. In a previous study \cite{TC}, a different aspect of ribosome recyling was considered. This work focused on the enhancement of the ribosome concentration at the initiation site, as a result of diffusion of the ribosome subunits from the termination site (of the same mRNA). Due to the spatial proximity of initiation and termination sites, $\alpha $ is effectively increased. By contrast, our investigation of constrained resources leads to an effective \emph{reduction} of $\alpha $.

In the context of traffic models, our problem corresponds to a
generalization of the parking garage problem \cite{Ha}. Here, one special
site (the ``parking garage'', or reservoir) is introduced into a TASEP on a
ring (lattice with periodic boundary conditions) and, for this site only,
the occupancy is unlimited. Particles (``cars'') jump into the garage with
unit rate (corresponding to $\beta =1$), irrespective of its occupancy.
Particles exit the garage with rate $\alpha $, provided the site following
the garage is empty. As we will see presently, this is a special case of our
more general model.

We begin with a description of our model and its observables of interest.
Next, we outline results from simulations and compare them to a simple
theory which builds on exact analytic results for the standard TASEP. We
find excellent agreement for nearly all ($\alpha $,$\beta $) and conclude
with some comments and open questions.

\section{The model}

The standard TASEP is defined on a one-dimensional lattice of length $L$,
with sites labeled by $i$, $i=0,1,...,L-1$. Each site can be empty or
occupied by a single particle, reflected through a set of occupation numbers
$\{n_i\}$ which take the values $0$ or $1$. The boundaries of the lattice
are open. Particles hop from a reservoir onto the lattice with a rate $%
\alpha $. Once on the lattice, particles will hop to the nearest-neighbor
site on the right, provided it is empty. Once a particle reaches the end of
the lattice, it hops back into the reservoir with rate $\beta $. The
dynamics is random sequential, leading to fluctuations in the local
occupations as well as in the total number of particles, defined via

\[
N\equiv \sum_{i=0}^{L-1}n_{i}
\]
The overall density is given by $\rho \equiv N/L${. Ensemble averages are
denoted by }$\left\langle ...\right\rangle ${. }

\begin{figure}[tbp]
\par
\begin{center}
\includegraphics[height=0.45\textwidth]{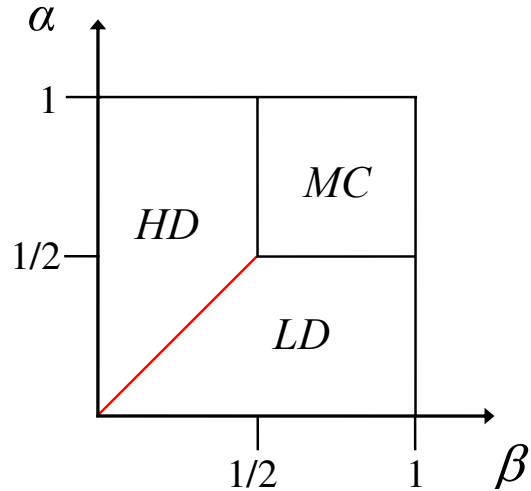}
\end{center}
\par
\vspace{-2mm}
\caption{Phase diagram of the standard TASEP.}
\label{fig:phase_diagram}
\end{figure}

Three distinct phases can be identified and are displayed in an $\left(
\alpha ,\beta \right) $ phase diagram (see Fig. \ref{fig:phase_diagram}): a
low density (LD), a high density (HD), and a maximum current (MC) phase. The
phases are distinguished, respectively, by their average densities: $\rho
=\alpha \equiv \rho _{-}$, $\rho =1-\beta \equiv \rho _{+}$, and $\rho
_{MC}=1/2$. The corresponding stationary currents are given by $J_{-}\equiv
\alpha (1-\alpha )$, $J_{+}\equiv \beta (1-\beta )$, and $J_{MC}\equiv 1/4$.
In all of these expressions, finite-size corrections have been neglected.
Because of particle-hole symmetry, all aspects of the HD and LD phases are
related. The phase boundaries between the HD and MC phases and between the
LD and MC phases mark continuous transitions. The line separating the HD and
LD phases, $\alpha =\beta <1/2$, is a coexistence line. Here, the system
consists of a region with low density ($\rho _{-}$) followed by one with
high density ($\rho _{+}$), connected by a microscopically sharp ``shock.''
This shock diffuses freely between the ends of the system, so that the \emph{%
average} density profile is linear. Often this is referred to as the ``shock
phase'' (SP).

The system of interest here differs from the standard TASEP in one important
way: The number of particles in the reservoir, $N_{p}$, is finite, and we
choose the on-rate $\alpha $ to depend on $N_{p}$ as follows. So as to
distinguish this varying rate from the $\alpha $ of the ordinary TASEP, we
denote the former as the \emph{effective} on-rate $\alpha _{eff}$ and write
\[
\alpha _{eff}\left( N_{p}\right) =\alpha f(N_{p})\,\,,
\]
where $f(x)$ is a function that satisfies three conditions: (i) $f(0)=0$,
(ii)\ $f(\infty )=1$, and (iii) $f(x)$ is monotonically increasing. The
first of these conditions is self-evident; the second simply connects our
model to the standard one with unlimited resources; the last is just common
sense. In this notation, the ``parking garage'' problem \cite{Ha} is
characterized by $f(N_{p})=\theta (N_{p})$ where $\theta $ is the Heaviside
step function. Here, we wish to model, say, a cell with finite number of ribosomes, so that it is natural to assume that a smaller number of
particles in the pool results in a lower on-rate, etc., and that $f$ would
be a smoother function. Specifically, for the simulations shown below we
choose
\begin{equation}
f(N_{p})=\tanh \left( N_{p}/N^{\ast }\right)  \label{alpha-MCS}
\end{equation}
where $N^{\ast }$ provides the scale for crossover to saturation. The
detailed choice of $N^{\ast }$ is unimportant, but it is convenient to make
it extensive in $L$. For easier comparison, we choose $N^{\ast }$ to be the
average (stationary) density of the unconstrained TASEP defined by the
parameters $\left( \alpha ,\beta \right) $: For example, if $\alpha
>1/2,\beta <1/2$ (HD phase, ordinarily), we set $N^{\ast }$ to $L\rho
_{+}=L(1-\beta )$. For the SP, we arbitrarily choose $N^{\ast }=L/2$. To
summarize, the control parameters of our model are the same as the ordinary
TASEP ($\alpha ,\beta ;L$) plus the total number of particles, $N_{tot}$
(i.e., $N+N_{p}$).

To characterize our system in steady state, we measure the average density $%
\rho \equiv \langle N\rangle /L$, the average (local) density profile $\rho
_i\equiv \langle n_i\rangle /L$, and the average current $J$. For the last
quantity, we record the total number of particles entering and leaving the
chain over the run (thereby improving statistics) and divide the average of
these by the length of the run.

In our Monte Carlo simulations, we pick randomly from the $N$ particles on
the lattice and one additional \emph{virtual }particle. If a lattice
particle is selected, we attempt to update its position as in ordinary
TASEP; if the virtual particle is selected, we attempt to place a new
particle on the first site of the chain. $N+1$ attempts constitute one Monte
Carlo Step (MCS). Initially, the chain is empty, i.e., $N=0$, $N_p=N_{tot}$.
Typically, the first $10^6$ ($10^7$ for SP) MCS are discarded, to ensure
that the system has reached steady state, before data are taken. Then,
configurational observables are recorded every $100$ MCS and averaged until
the run terminates. The typical length of each run is $10^7$ ($10^8$ for SP)
MCS. Longer runs are required to obtain low noise data for density profiles
and currents. Statistical errors were estimated through visual inspections
of time series and through multiple runs, to ensure the reproducibility of
the data. We simulated system sizes in the range of $L=250$ to $L=8000$,
with most data taken for $L=1000$. To obtain the phase diagram, we simulated
more than a dozen $\left( \alpha ,\beta \right) $ pairs. Here, we report
results for four points: $\left( 0.25,0.75\right) $, $\left(
0.75,0.25\right) $, $\left( 0.75,0.75\right) $, and $\left( 0.25,0.25\right)
$. Since these correspond, respectively, to LD, HD, MC, and SP in a standard
TASEP, we will use these letters to refer to the four cases studied.

\section{Results from simulations}

In this section, we summarize our simulation results. Since only the on-rate
depends on the feedback, the behavior of our system is no longer governed by
particle-hole symmetry. As a result, the simplest case is the first in the
above, i.e., ``LD''. Both $\rho $ and $J$ increase with $N_{tot}$ in an
expected fashion. The next case, MC, is already slightly more interesting:\
Both $\rho $ and $J$ increase until the effective on-rate reaches 1/2, after
which both become constant. The end result is a pronounced kink in $\rho $.
The HD case provides an even more interesting scenario: The system
properties range through \emph{three} regions as $N_{tot}$ is increased.
Finally, the SP provides the most unexpected behavior. In the next section,
we turn to a simple theoretical description of our findings.

\emph{The LD case.} Here, we set $\alpha <1/2$, $\beta >\alpha $, and vary
the total number of particles, $N_{tot}$, in the system. With $f$ given by
Eqn (\ref{alpha-MCS}), our simulation results for $L=1000$ and $\left(
\alpha ,\beta \right) =\left( 0.25,0.75\right) $ are shown in Fig. \ref%
{fig:LD}. As expected, we see that, for large $N_{tot}$, the system density $%
\rho =\left\langle N\right\rangle /L$ approaches the value for the standard
TASEP for $N_{tot}\gg N^{\ast }$ (here, $0.75L$), namely, $\rho =\alpha $
(which is $0.25$ in our case). More interestingly, for $N_{tot}\lesssim
N^{\ast }$, we find a reduced density, $\rho <\alpha $, a signal that the
system is responding to the limited availability of particles
from the reservoir. In the limit $N_{tot}\rightarrow 0$, $\rho $ naturally
vanishes, with a predictable slope (see below). The current $J$ in this
``phase'' also decreases monotonically with decreasing $N_{tot}$, from its
asymptotic limit $\beta (1-\beta )$ (given by $0.1875$).

\begin{figure}[tbh]
\begin{center}
\includegraphics[height=0.35\textwidth]{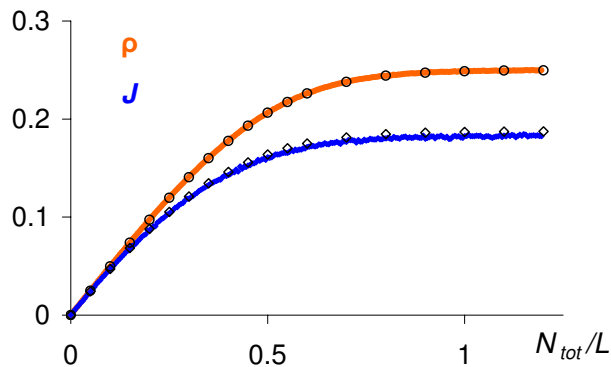}
\end{center}
\caption{Average density $\protect\rho $ and current $J$ vs $N_{tot}/L$ for
an LD case ($L=1000$, $\protect\alpha =0.25$, and $\protect\beta =0.75$).
Open circles and diamonds are from the analysis in Section 4.}
\label{fig:LD}
\end{figure}

\begin{figure}[tbh]
\par
\begin{center}
\includegraphics[height=.35\textwidth]{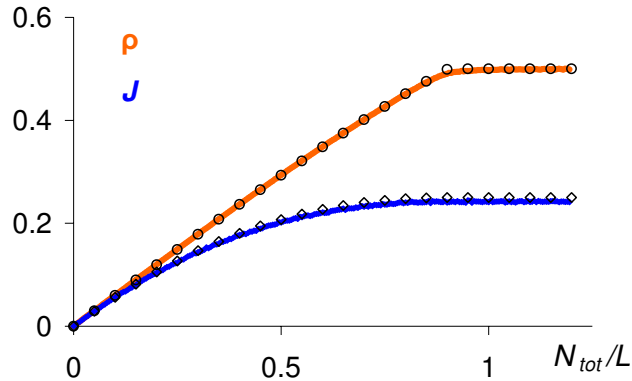}
\end{center}
\caption{Average density $\protect\rho $ and current $J$ vs $N_{tot}/L$ for
an MC case ($L=1000$, $\protect\alpha =0.75$, and $\protect\beta =0.75$).
Open circles and diamonds are from the analysis in Section 4.}
\label{fig:MC}
\end{figure}

\emph{The MC case.} This domain of the phase diagram is characterized by $%
\alpha >1/2$ and $\beta >1/2$ in the standard TASEP. Fig. \ref{fig:MC} shows
the averages $\rho $ and $J$, as $N_{tot}$ varies, for $(\alpha ,\beta
)=(0.75,0.75)$. The former displays a single kink, just below $%
N_{tot}/L\simeq 1.0$, accompanied by a rather smooth crossover in the
current. As we will see below, this kink is associated with a crossover from
an MC-like behavior to an LD-like behavior, as $N_{tot}$ drops below the
value required to sustain a density $\rho _{MC}=1/2$ on the chain.

\begin{figure}[tbh]
\begin{center}
\includegraphics[height=.35\textwidth]{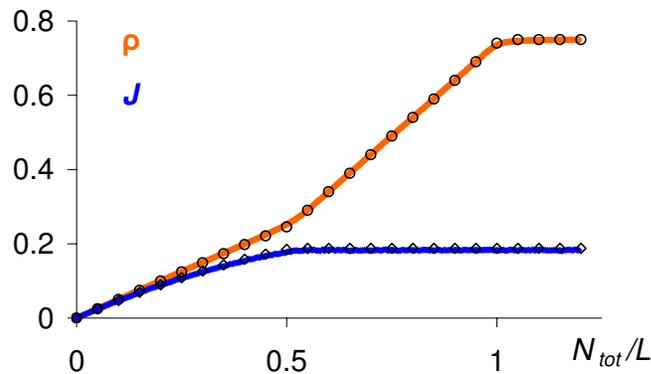}
\end{center}
\caption{Average density $\protect\rho $ and current $J$ vs $N_{tot}/L$ for
an HD case ($L=1000$, $\protect\alpha =0.75$, and $\protect\beta =0.25$).
Open circles and diamonds are from the analysis in Section 4. The
predictions from domain wall theory (not shown) are similiar.}
\label{fig:HD}
\end{figure}

\begin{figure}[tbh]
\begin{center}
\includegraphics[height=.35\textwidth]{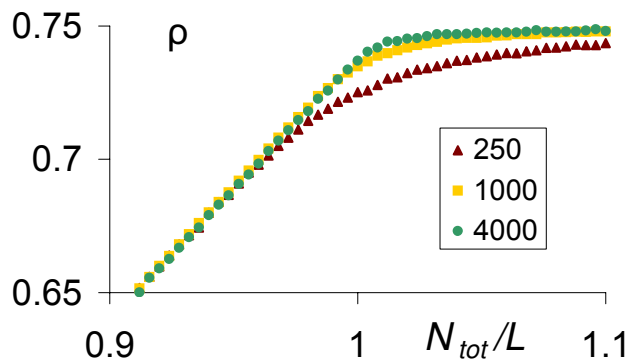}
\end{center}
\caption{Finite-size effects on the kink in $\protect\rho $, near $%
N_{tot}/L=1.0$, for the HD case in Fig. 4:$\;L=250$, $1000$, and $4000$. }
\label{fig:HD_Zoom}
\end{figure}

\emph{The HD case. }Here, we set $\beta <1/2$ and $\alpha >\beta $. Fig. \ref%
{fig:HD} shows our results for $(\alpha ,\beta )=\left( 0.75,0.25\right) $.
As expected, for sufficiently high $N_{tot}$, the system settles into the
density and current associated with the HD phase of the standard TASEP: $%
\rho =0.749\simeq \rho _{+}$, and $J=0.184\simeq \beta \left( 1-\beta
\right) $. As $N_{tot}$ is reduced, to the extent that the chain is
prevented from sustaining a density of $\rho _{+}$, the naive expectation is
that a crossover to an LD-like behavior appears. Remarkably, however, Fig. %
\ref{fig:HD} shows not just two, but \emph{three} distinct regimes,
separated by two pronounced ``kinks'' at $N_{tot}\,/L\simeq 0.5$ and $1.0$%
. As Fig. \ref{fig:HD_Zoom} illustrates, these kinks become sharper with
increasing $L$. Looking for potentially critical behavior, we measured the
variance of $\langle N\rangle $ near the kinks and compared it to its values
deep within each regime. To our surprise, at least for the $L=1000$ system,
the variance was not noticeably larger at the kinks than at other locations.
Simulations at larger values of $L$ would be required to settle this issue
with more certainty.

The current $J$ is also displayed in Fig. \ref{fig:HD}. For low values of $%
N_{tot}/L$, $J$ increases in an expected way. Around $N_{tot}=0.5L$, it
shows a distinct kink and then becomes essentially constant, with a value of
$0.184$. This is close to $0.1875$, the value of $J$ for infinite $L$. Since
the current appears to have already reached the ``asymptotic'' value here,
it is not too surprising that there is no sign of a ``second kink'' (near $%
N_{tot}=L$, where $\rho $ has another one). Below, we offer a theoretical
description for these findings.

\begin{figure}[tbh]
\begin{center}
\includegraphics[height=.4\textwidth]{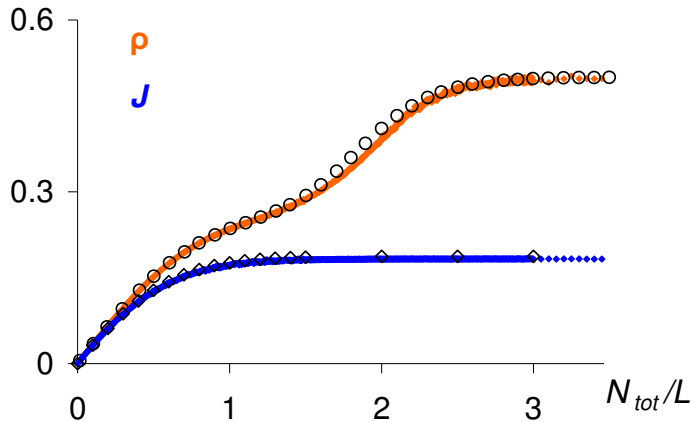}
\end{center}
\caption{Average density $\protect\rho $ and current $J$ vs $N_{tot}/L$ for
an SP case ($L=1000$, $\protect\alpha =0.25$, and $\protect\beta =0.25$).
Open circles are from the domain wall theory in Section 4. Open diamonds are
the current computed from Eqn (\ref{LD-th}). }
\label{fig:SP}
\end{figure}

\begin{figure}[tbh]
\begin{center}
\includegraphics[height=.40\textwidth]{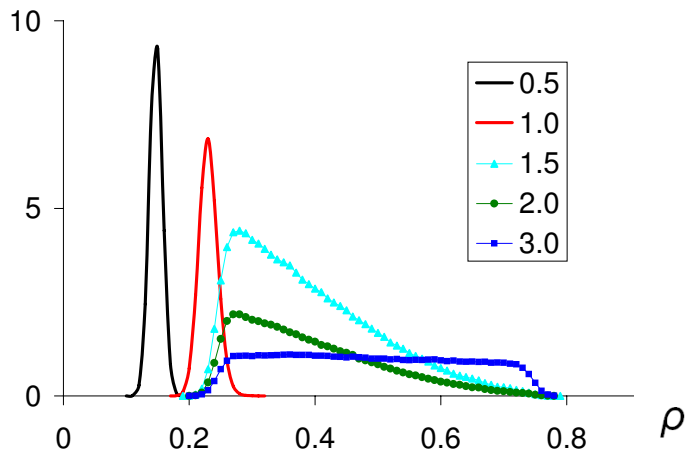}
\end{center}
\caption{Histograms of the density $\protect\rho $ for $L=1000$, $\protect%
\alpha =0.25$, and $\protect\beta =0.25$. (For ease of comparisons, the
histograms with solid lines only are scaled down by 2.) Legend shows various
values of $N_{tot}/L$.}
\label{fig:SP-hist}
\end{figure}

\emph{The SP case. }Defined in the standard TASEP by $\alpha =\beta <1/2$,
this line of first-order phase transitions separates the HD and LD phases.
Though the \emph{average }density is 1/2, its fluctuations are large. In
particular, since a shock (which separates a region with density $\rho _{+}$
from one with $\rho _{-}$) diffuses freely between the ends of the system,
the probability $P(\rho )$ of finding the system with a density $\rho $
approaches a flat distribution: $P(\rho )=\theta (\rho -\rho _{-})\theta
(\rho _{+}-\rho )/\left( \rho _{+}-\rho _{-}\right) $, in the limit $%
L\rightarrow \infty $. Of course, finite-size effects will smooth out the
steps at $\rho _{+}$ and $\rho _{-}$. Imposing finite resources, we may
expect that $\rho $ increases simply with $N_{tot}$ as in the LD case,
perhaps rising more slowly to the asymptotic value of 1/2. Instead, its
properties are much more subtle. As illustrated in Fig. \ref{fig:SP} (for $%
\alpha =\beta =0.25$), $\rho $ increases with $N_{tot}/L$\textbf{,} slows,
speeds up, and slows down again. The behavior in the low $N_{tot}$ regime is
intuitively understandable, since there are too few particles to sustain any
phase other than LD. The large $N_{tot}$ regime is also expected. In the
crossover regime ($L\lesssim N_{tot}\lesssim 3L$), an intriguing
``shoulder'' develops. Though this behavior is accounted for by the domain
wall theory \cite{DW,SA} (see below), we have found no simple and intuitive
way to understand it. Perhaps the presence of large fluctuations, as the
system gets closer to the SP, smooths out the two kinks observed in the HD
case. To gain more
insight, we investigated $P(\rho )$. Fig. \ref{fig:SP-hist} shows particle
density histograms (gathered from 50K data points and plotted with arbitrary
normalization here), for several choices of $N_{tot}/L$. For low $N_{tot}/L$
(e.g., 0.5 in the figure), our system is dominated by the lack of resources,
so that this distribution is essentially Gaussian, quite similar to those
measured deep within the LD phase. As $N_{tot}/L$ increases, the system
enters the crossover regime and $P(\rho )$ becomes quite asymmetric, as a
detailed fitting of the $N_{tot}/L=1.0$ case shows. At the other extreme,
this distribution is indeed relatively flat. However, the true asymptotic
regime is reached only for very large $N_{tot}/L$. Thus, the slope for even
the $N_{tot}/L=3.0$ case (cf. Fig. \ref{fig:SP-hist}) is discernably
non-zero. Throughout the intermediate regime, the structure of $P$ is more
interesting. Each histogram has the cross-section of a ``lean-to'' (a shack
with a slanted roof). A more detailed inspection shows that the ``roof'' is
predominantly a (slowly) decaying exponential \footnote{%
Such distributions are also found in unconstrained TASEP's with $\alpha $
nearly identical to $\beta $.}. This picture is consistent with the one from
domain wall theory \cite{DW,SA}, as will be discussed below. It is this slow
crossover - from well-defined Gaussians to a completely flat distribution -
which accounts for the interesting structure in the rise of $\rho $ with $%
N_{tot}/L$.

Finally, the behavior of the current is also sensitive to the slow
crossover, although it displays no interesting structures like those in $%
\rho $. Comparing Fig. \ref{fig:SP} with the earlier ones, we see that it
saturates at values of $N_{tot}/L$ which are 2 to 3 times higher than in the
other cases.


\section{Analytic predictions}

In the steady state of a standard TASEP, the overall density of particles is
controlled only by the entrance and exit rates. It is natural to expect that
these relations would also hold in our case, except that the stationary
density needs to be determined from the (variable)\ entrance rate in a
self-consistent way. In the following, we pursue this approach and compare
it to the simulation results reported in the preceding section.

\emph{The LD case.} Again, we begin with a choice of $(\alpha ,\beta )$ such
that the system is in an LD phase, with $\rho <1/2$. Invoking the exact
expression for the density of the standard TASEP, namely, $\rho =\alpha $,
we write the density of particles in the resource-limited case as $\rho
=\alpha _{eff}=\alpha f(N_p)$. With $N_p=N_{tot}-L\rho $, this becomes an
implicit equation for $\rho $:
\begin{equation}
\rho =\alpha f(N_{tot}-L\rho )  \label{LD-th}
\end{equation}
which can be solved for given $\alpha $ and $N_{tot}$. Specifically, since $%
0\leq f(N_{tot}-L\rho )\leq 1$, the solution will lie between $0$ and $%
\alpha $. The upper bound will be reached if $N_{tot}/L\rightarrow \infty $
so that the right hand side of Eqn (\ref{LD-th}) remains essentially
constant as a function of $\rho $. For any finite value of $N_{tot}/L$, the
right hand side is a monotonically decreasing function of $\rho $ so that
(i) Eqn (\ref{LD-th}) always has a solution, and (ii) the solution decreases
monotonically with $N_{tot}/L$. The resulting density, and the associated
current, $J=\alpha _{eff}\left( 1-\alpha _{eff}\right) $, are shown as open
circles and diamonds, respectively, in Fig. \ref{fig:LD}, for the choice of $%
f$ given in Eqn (\ref{alpha-MCS}). We see, in particular, that for small
values of $N_{tot}/L$, we may use $\tanh x\simeq x$ whence $\rho \simeq
N_{tot}/(2L)$. Clearly, the agreement with the simulation data is impressive.

\emph{The MC case.} After these considerations, the analysis of the MC phase
is quite straightforward. For large $N_{tot}$, the density saturates at $%
\rho =1/2$, with $J=1/4$. For small $N_{tot}$, the system is LD-like and
follows Eqn (\ref{LD-th}). The transition occurs when $\alpha _{eff}$ is
forced to drop below $1/2$, i.e., at the intersection of Eqn (\ref{LD-th})
and $\rho =1/2$. Here, $N_{tot}$ assumes the critical value of $N_{tot,c}$,
given by
\begin{equation}
N_{tot,c}=L\rho +f^{-1}\left( \frac{1}{2\alpha }\right) \,.  \label{MC-th}
\end{equation}
The current follows the behavior of the density.

\emph{The HD case.} Next, we turn to the high-density phase, with $\beta
<1/2 $ and $\alpha >\beta $. If the number of available particles is very
large, we may again invoke the exact relation for the density of the
standard TASEP, namely
\begin{equation}
\rho =\rho _{+}=1-\beta \,.  \label{HD-th1}
\end{equation}
This reflects the constant density observed in the simulations. Clearly,
this cannot be sustained as the number of available particles is reduced,
since the effective $\alpha $ of the system is also reduced. A transition
should be expected when it reaches $\beta $. $=\alpha _{eff}=\alpha f(N_p)$.
Translating $\beta $ $=\alpha _{eff}$ into an equation for the system
density $\rho $, we arrive at
\begin{equation}
\beta =\alpha _{eff}=\alpha f(N_{tot}-L\rho )\,\,.  \label{HD-th1a}
\end{equation}
Solving for $\rho $, and setting it to the one from Eqn (\ref{HD-th1}), we
find the critical value of the total particle number for this transition: $%
N_{tot.2}=L\left( 1-\beta \right) +f^{-1}\left( \beta /\alpha \right) $.
(The subscript, $2$, helps to distinguish it from the lower transition point.)
For the specific $f$ used in simulations, this results in
\begin{equation}
N_{tot.2}=L\left( 1-\beta \right) \left[ 1+\tanh ^{-1}\frac \beta \alpha %
\right] .  \label{HD-c2}
\end{equation}
As we see, $N_{tot.2}/L=0.75\left[ 1+\tanh ^{-1}\left( 1/3\right) \right]
\simeq 1.01$ is in excellent agreement with the ``second kink'' shown in
Fig. \ref{fig:HD}.

At the opposite extreme, the resources are very low, so that $\alpha _{eff}$
is significantly less than $\beta $, and the system is pushed into LD-like
behavior. For such low $N_{tot}$, the density follows Eqn (\ref{LD-th}). In
this regime, $f$ is well approximated by a linear function, so that $\rho
\simeq \alpha (N_{tot}-L\rho )/N^{\ast }=\alpha (N_{tot}/L-\rho )/\left(
1-\beta \right) $, i.e.,
\[
\rho \simeq \frac{\alpha }{1-\beta +\alpha }N_{tot}/L\,\,.
\]
Of course, Eqn (\ref{LD-th}) is also not ``sustainable,'' in the sense that $%
\rho \rightarrow \alpha $ for large $N_{tot}$ instead of the correct limit: $%
\rho \rightarrow 1-\beta $. The critical value of $N_{tot}$ again occurs
when $\alpha _{eff}$ reaches $\beta $, i.e., when the density given by Eqn (%
\ref{LD-th}) reaches $\beta $. The result is
\begin{equation}
N_{tot.1}=L\beta +L\left( 1-\beta \right) \tanh ^{-1}\left( \frac{\beta }{%
\alpha }\right) .  \label{HD-c1}
\end{equation}
Again, for the case shown in Fig. \ref{fig:HD}, there is excellent agreement
between the data and this prediction: $N_{tot.1}/L\simeq 0.51$.

Between $N_{tot.1}$ and $N_{tot.2}$, Eqn (\ref{HD-th1a}) provides a \emph{%
linear} dependence of $\rho $ on $N_{tot}/L$
\begin{equation}
\rho =N_{tot}/L-f^{-1}\left( \beta /\alpha \right) /L\,\,,  \label{HD-th2}
\end{equation}
with \emph{unit }slope. Again, with \emph{no} fitting parameters, this
theoretical prediction agrees remarkably well with the observations. In this
regime, the reservoir occupation (and so $\alpha _{eff}$) remains constant,
while the change in $N_{tot}$ is balanced by the change in total lattice
population: $L\rho $. Since $\alpha _{eff}=\beta $ (on the average), the
current remains at the \emph{constant} value $J=\beta \left( 1-\beta \right)
$ throughout.

While this ``three-piece'' approach is both intuitively appealing and
surprisingly successful in reproducing simulation data, it begs the
question: Is there a unified theory that is just as successful with
predictions? The answer is affirmative, namely, the domain wall theory \cite%
{DW,SA}. Since this theory is absolutely indispensable for providing good
predictions for the next case, we will discuss its details in the next
subsection.

\emph{The SP case. }Here, Monte Carlo simulations presented us with
unexpected phenomena: the ``shoulder'' in Fig. \ref{fig:SP} for $\rho $. The
simple approach taken above succeeds only in predicting the behavior for the
extreme values of $N_{tot}/L$. Dominated by the limited resources, the
solution of $\rho =\alpha _{eff}$ fits the data very well for $%
N_{tot}/L\lesssim 1$. At the other extreme ($N_{tot}/L\gg 1$), the system is
necessarily in the SP, where $\rho =1/2$ on the average. In the crossover
regime, the understanding of the structure in $\rho \left( N_{tot}/L\right) $
requires a more sophisticated theory \cite{DW,SA}. Based on a biased random
walk of the shock (i.e., domain wall), Santen and Appert \cite{SA} derived
the average occupation for a \emph{finite} TASEP:
\begin{equation}
\left\langle N\right\rangle =\rho _{-}L+\delta \left\{ \frac r{1-r}-\left(
L+1\right) \frac{r^{L+1}}{1-r^{L+1}}\right\}  \label{<N>}
\end{equation}
where $\delta =\rho _{+}-\rho _{-}=\left( 1-\beta \right) -\alpha $, and
\begin{equation}
r=\frac{\alpha \left( 1-\alpha \right) }{\beta \left( 1-\beta \right) }
\end{equation}
is the ratio of the diffusion constants to the right/left. To apply to our
SP case, we simply set $\alpha $ to be $\alpha _{eff}=\beta f(N_{tot}-L\rho
) $ in this equation. That sets up a relationship between $\rho $ and $%
N_{tot}/L$ which can be computed numerically. The agreement between this
theory and data is quite respectable (Fig. \ref{fig:SP}), though not as
spectacular as in the other cases. Not surprisingly, most of the
disagreement occurs in the crossover region. We believe that the main
difficulty lies with the large fluctuations of the domain wall, feeding back
in non-trivial ways to the on-rate.

For another perspective on the successes and limitations of this theory, we
can compare $P(\rho )$ above with the predicted $P_{i}$, the probability of
finding the domain wall to be at site $i$. The latter is, for the steady
state, a pure exponential \cite{DW,SA}: $P_{i}\propto \exp \left( i/\xi
\right) $. Since our focus here is the dependence on $\rho $, $P_{i}$
can be translated into $P(\rho )$, via the relation $\rho =\rho _{+}\left(
1-i/L\right) +\rho _{-}\left( i/L\right) $. We find that the \emph{%
predicted} $P(\rho )$ is proportional to
$\exp \left( -\rho /\sigma \right) $, and nonzero only in the interval
$[\rho _{-},\rho _{+}]$. Our
simulations show several non-trivial deviations from this theoretical
prediction, however. (i) The slopes in the exponent are systematically
steeper than predicted. (ii) Quadratic terms in $\ln P(\rho )$ are far from
negligible. (iii) The tails of $P(\rho )$ extend substantially outside the
range $\left[ \rho _{-},\rho _{+}\right] $. In other words, there is room
for improvements to the domain wall theory.

Another intriguing phenomenon, not shown here explicitly, is the complex
dependence of this crossover on $L$. In particular, both simulations and
theory show that the ``shoulder'' completely disappears for small $L$. For
example, in an $L=100$ run, $\rho \left( N_{tot}/L\right) $ rises linearly
for $N_{tot}/L\lesssim 1.5$ and simply ``slows down'' into the asymptotic
value of $1/2$ for $N_{tot}/L\gtrsim 2.5$. For large $L$, in contrast, there
exists a subtle competition between the factors $\left( L+1\right) $ and
$r^{L+1}$ in Eqn (\ref{<N>}). The result is that the crossover region gets
longer with larger $L$, producing challenges at the simulations front as
well. Of course, these difficulties are hardly surprising, since the
complexities here are undoubtedly due to large fluctuations associated with
the infrared scales. Further studies into this issue, especially more
detailed investigations of the lean-to's in $P(\rho )$, should provide
insight for a better understanding of such fluctuations.

To conclude this section, we briefly note that we also simulated an
alternate form of the on-rate, namely,
$f(N_{p})= 1/\left[ 1+\left(N^{\ast }/N_{p}\right) ^{2}\right] $.
We will not present any details here,
but the agreement of simulation data and analytic results for the three
regimes (HD, LD, and MC)\ is equally convincing. However, the details of the
crossover regime in SP are likely to be quite sensitive to functional form
of $f$.

\section{Conclusions}

In this paper, we studied a TASEP with open boundary conditions. In contrast
to the standard model, where the particles are supplied by an infinite
reservoir, the total number of available particles in our model, $N_{tot}$,
is \emph{fixed}. Mimicking limited resources, the available particles are
shared between the reservoir and the system itself: If more particles bind
to the chain, the reservoir is depleted, and vice versa. We account for this
situation through an effective on-rate, $\alpha _{eff}$, which depends on
$N_{p}$, the number of particles in the pool, and hence, on $N$, the number
on the chain. Specifically, we mainly used Eqn (\ref{alpha-MCS}), $\alpha
_{eff}=\alpha \tanh \left( N_{p}/N^{\ast }\right) $, here. As the number of
available particles is reduced from infinity, the system crosses over from
the asymptotic behavior of the standard TASEP to different behaviors
dominated by limited resources.

If the corresponding standard model (i.e., $N_{tot}\rightarrow \infty $
limit) lies in the LD phase, there are no surprises in either $\rho $, the
average particle density on the chain, or $J$, the average current: Both are
proportional to $N_{tot}$ for $N_{tot}\ll L$, crossing gently over to the
asymptotic values much like the $\tanh $ function. For the MC phase, there
is already a sublety:\ both $\rho $ and $J$ increase monotonically and then,
at a characteristic $\alpha _{eff}$ ($=1/2$) become constant. $\rho $
displays a notable kink there. For the other two cases (HD and SP), $\rho $
exhibits an additional, intermediate regime (though the behavior of $J$
shows little hint of this regime). In the HD case, $\rho $ is linear in
$N_{tot}/L$ with \emph{unit }slope; only the intercept depends on $\alpha
/\beta $. Here, $N_{p}$ remains constant, as any increase in $N_{tot}/L$
is ``absorbed'' by the chain (through a shift in the domain wall). This
behavior may be regarded in the same light as an equilibrium system with
phase co-existence. Considering, e.g., the pressure \emph{vs.} volume
isotherm for a liquid-gas system below criticality, the pressure remains
constant over some range, as any increase in volume is absorbed by a shift
in the liquid-to-gas ratio.

The most interesting and challenging case is the SP. In the intermediate
regime, a ``shoulder'' appears in $\rho \left( N_{tot}/L\right) $. In
addition, this structure displays a non-trivial dependence on $L$. Although
the predictions of domain wall theory agree reasonably well with MC data,
there are small ($\lesssim 6\%$) systematic discrepancies. We also
investigated histograms of particle densities, which cross over from sharply
peaked Gaussians for $N_{tot}\ll L$ to a flat distribution (between $\rho
_{-}=\beta $ and $\rho _{+}=1-\beta $) for $N_{tot}\gg L$.
In the intermediate regime, this distribution displays a much richer behavior
than a truncated exponential -- the result of domain wall theory.

Several questions remain open for further study. First, are the transitions
between the three regimes in HD associated with true thermodynamic
singularities? We find that the changes in slope accompanying these
``crossovers'' become sharper with increasing chain length $L$. While this
speaks for a true transition, the absence of large fluctuations is somewhat
surprising. Clearly, a more careful analysis is required to explore these
questions more fully.

Second, though domain wall theory is reasonably successful for $\rho \left(
N_{tot}/L\right) $ in the SP case, we would like to find improvements of
this approach, so as to account for the crossover regime better. Arriving at
a good understanding of $P(\rho )$ and the large fluctuations would be
desirable. Progress along these lines may also help us to develop an
intuitive picture for the complex $L$ dependence.

Finally, we are mindful of one of the motivations of this study, namely,
finite resources in biological systems, e.g., ribosomes (and aa-tRNA's) for
modeling protein synthesis in a cell \cite{MG,MG2,LBSZia,TomChou,JJ,JJ2}.
For that application, the above TASEP needs to be generalized, to include
exclusion at a distance (particles covering more than one site) and
inhomogeneous hopping rates. How these systems are affected by finite
resources will pose many interesting new challenges, especially on the
theoretical frontier.

\emph{Acknowledgements. }We thank Travis Merritt for preliminary simulation
results and Meesoon Ha and Marcel den Nijs for helpful discussions. This
work was supported in part by the National Science Foundation through
DMR-0414122 and DMR-0705152.

\end{document}